\documentclass[twocolumn,twoside,slac_two]{revtex4}
\usepackage{graphicx}
\usepackage{fancyhdr}
\usepackage{overpic}
\usepackage{relsize}
\usepackage{multirow}
\usepackage{booktabs}
\usepackage{xspace}
\usepackage{ifthen}

\newif\ifflatbib
\providecommand{\flattenBib}{false}
\ifthenelse{\equal{\flattenBib}{true}}{
  \flatbibtrue
}{
  \flatbibfalse
}

\makeatletter
\DeclareRobustCommand\bfseries{%
  \not@math@alphabet\bfseries\mathbf
  \fontseries\bfdefault\selectfont\boldmath}

\makeatother

\bibpunct{[}{]}{,}{n}{}{,}

\graphicspath{{figures/}}

\newcommand{\babar}{\mbox{%
    \slshape B\kern-0.1em{\smaller A}\kern-0.1em
    B\kern-0.1em{\smaller A\kern-0.2em R}}\xspace}

\newcommand{\invfb}{\ensuremath{\,\text{fb}^{-1}}\xspace}

\newcommand{\EE}[1]{\ensuremath{\cdot 10^{#1}}}

\newcommand{\prelim}{\ensuremath{^*}}

\def\BR{{\ensuremath{\cal B}}\xspace}

\def\piz   {\ensuremath{\pi^0}\xspace}

\def\Kbar  {\kern 0.2em\overline{\kern -0.2em K}{}\xspace}

\def\Kz    {\ensuremath{K^0}\xspace}
\def\Kzb   {\ensuremath{\Kbar^0}\xspace}
\def\KzKzb {\ensuremath{\Kz \kern -0.16em \Kzb}\xspace}
\def\nut        {\ensuremath{\nu_\tau}\xspace}

\newcommand{\tauknu}   {\ensuremath{ \tau^{-} \to K^{-} \nut}\xspace}
\newcommand{\BFtautoknu}    {\ensuremath{\BR(\tauknu)}\xspace}                  

\begin{document}

\title{Tau decays at the B-factories}

%

\author{A.~Lusiani}
\affiliation{Scuola Normale Superiore and INFN, Pisa}

\begin{abstract}
Since 1999, the B-factories collaborations \babar and Belle
have accumulated and studied large samples of tau lepton pairs. We
summarize in the following the most interesting recent results.
\end{abstract}

\maketitle

\thispagestyle{fancy}


\section{Introduction}
Today progress in tau physics proceeds mostly at the two B-factories
\babar and Belle, where unprecedented statistics of tau pair events
produced in $e^+e^-$ annihilations have been recorded. 

Both experiments rely on ``B-factories'' operating at a centre-of-mass
energy at and around  the $\Upsilon(4s)$ peak (10.58\,GeV).
\babar operates at the PEP-II complex at SLAC, which collides 9\,GeV
electrons against 3.1\,GeV positrons, and has recorded about 531\invfb
of data by May 2008. Belle operates at the KEKB B-factory in Japan,
which collides 8\,GeV electrons against 3.5\,GeV positrons, and has
recorded about 831\invfb of data by May 2008.

The \babar~\cite{Aubert:2001tu} and Belle~\cite{Abashian:2000cg} detectors
share several similarities and both include a silicon microvertex
detector, a drift chamber, a 1.5\,T solenoidal superconducting magnet,
an electromagnetic calorimeter based on Cesium Iodide crystals, and a
segmented muon detector in the magnet return yoke.  The two
experiments differ in the particle identification strategy: Belle uses
an aerogel threshold Cherenkov detector together with time-of-flight
and tracker dE/dx, whereas \babar relies on a ring-imaging Cerenkov
detector supplemented by the dE/dx in the trackers.

With a total data set now exceeding 1.1\,ab$^{-1}$ of integrated
luminosity and a $e^+e^-\to\tau^+\tau^-$ cross-section at 10.58\,GeV
of 0.919\,nb~\cite{Banerjee:2007is}, B-factories recorded more than
$10^9$ tau pairs and contributed significant progress to tau
lepton physics.

\section{Searches for Lepton Flavor Violation}

\subsection{Common analysis features}

Most tau LFV decay searches at the B-factories look for low track
multiplicity events that have 1 against 1 or 3 tracks in the
center-of-mass system reference frame (CM).  The thrust axis is used
to define two hemispheres: one of them must be consistent with a tau
LFV decay, while the other one must be compatible with a known tau
decay. The LFV tau decay products are expected to match, within the
experimental resolution, both the tau mass and the expected tau
energy, i.e.\ half the CM energy. Analyses must account for the fact
that initial and final state radiation processes decrease the amount of
reconstructed energy, and that radiation in decay and Bremsstrahlung
from the decay products decrease both the reconstructed energy and the
reconstructed mass.  The energy is reconstructed with a typical
resolution of 50\,MeV and, when constraining the reconstructed energy
to half the CM energy, the invariant mass is reconstructed with a
resolution of about 10\,MeV.  LFV decays are detected by an excess of
events over the expected background within typically 2 or 3 standard
deviations of the expected energy and mass.

The amount of expected background is usually estimated assuming that
each background source has a mass and energy distribution as modeled
by the Monte Carlo simulation, and then obtaining the normalization of
all background sources by fitting the resulting total background
distribution to the data events in the sidebands of the signal region.
The signal efficiency is estimated
with a Monte Carlo simulation and usually lies between 2\% and 10\%
depending on the decay channel. Typical cumulative efficiency components
include 90\% for trigger, 70\% for geometrical acceptance and
reconstruction in the detector, 70\% for reconstructing the selected
track topology, 50\% for particle identification, 50\% for additional
selection requirements before checking the reconstructed energy and
mass, and 50\% for requiring consistency with the expected energy and
mass. The selection efficiency and background suppression are
optimized to give the best "expected upper limit" assuming that the
data contain no LFV signal. The optimization procedure and all
systematic studies are completed while remaining ``blind'' to data
events in the signal box in the energy-mass plane, in order to avoid
experimenter biases.

When the expected background in the signal region is of order one or
less, the number of signal events is normally set to the number of observed
events minus the background, while in presence of sizable background
the numbers of background and signal events are concurrently determined
from a fit to the mass distribution of events that have total energy
compatible with the expected one.

\begin{figure*}[t]
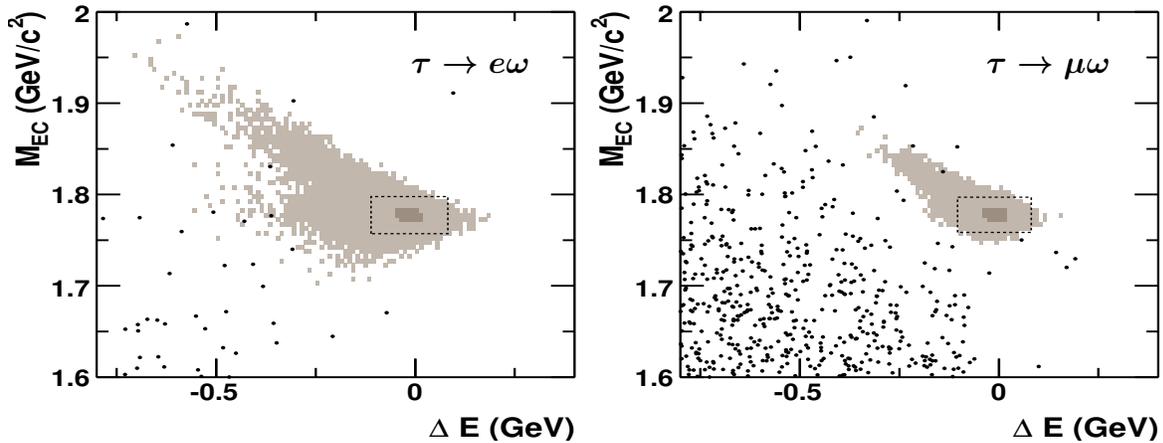

\centering
\begin{overpic}[width=0.9\linewidth,clip]{babarTauToEllOmega}
  \put(35,32){\larger\larger\textbf{$\tau\to e\omega$}}
  \put(85,32){\larger\larger\textbf{$\tau\to \mu\omega$}}
\end{overpic}
\caption{The dots represent data candidate events for the
  lepton-flavor-violating $\tau\to\ell\omega$ decay in the reconstructed
energy-mass plane. The gray areas indicate where 90\% of the signal
events would be expected according to the simulation. The rectangular
regions indicate where an excess of signal candidates is searched for.}
\label{tau-l-omega}
\end{figure*}

\subsection{Results}

LFV decays can be grouped in the following categories: tau to 
lepton-photon ($\tau\to\ell\gamma$, where $\ell=e,\mu$), tau to three
leptons or one lepton and two charged hadrons
($\tau\to\ell_1\ell_2\ell_3$, $\tau\to\ell h_1 h_2$), tau to a lepton
and a neutral hadron ($\tau\to\ell h^0$, where $h^0 = \pi^0, \eta,
\eta^\prime, K_s^0, \text{\textit{etc.}}$).


\babar has recently published results on a search for the
lepton-flavor-violating decays $\tau\to\ell \omega$ ($\ell=e,\mu$),
based on the analysis of $384\invfb$ of data\cite{Aubert:2007kx}.
Candidate $\omega$ mesons are reconstructed using the decay into
$\pi^+\pi^-\pi^0$; simulated signal events are selected with an
efficiency of 2.96\% (2.56\%) for the electron (muon) channel.
The signal is searched by requiring that the reconstructed tau
candidate has the correct mass ($M_{\text{RECO}}$) and the expected energy
($E_{\text{RECO}} \simeq \sqrt{s}/2$) in the CM.
The actual reconstructed quantities are smeared by the detector
resolution and 
shifted by radiation and Bremsstrahlung (see Figure~\ref{tau-l-omega}).
The amount of expected
background is estimated by determining the shape of the
two-dimensional probability
density function (PDF) of each background component on
simulated events or in control samples extracted from data for
background sources like Bhabha events whose simulation is practically
unfeasible. The normalization of all background component PDFs are
then determined with a fit on the data sidebands surrounding the
signal region in the $M_{\text{RECO}} - E_{\text{RECO}}$ plane.
No events are found while 0.35 (0.73) are expected for the electron
(muon) channel, and upper limits are determined 
according
to the Cousin and Highland prescription~\cite{Cousins:1991qz} with no
Feldman and Cousin ordering in the range as follows:
$\BR(\tau\to e\omega) < 11\EE{-8}$ and $\BR(\tau\to\mu\omega) <
10\EE{-8}$ at 90\% confidence level (CL).

Recently, \babar and Belle have published~\cite{Aubert:2007pw,Miyazaki:2007zw}
improved results on tau
to three leptons LFV searches, based on enlarged event samples of
$400\invfb$ and $535\invfb$ respectively.  The two analyses have
similar signal efficiencies ($5.5\%{-}12.5\%$ for the different
channels) although different
strategies are adopted to define the signal regions in the energy-mass
plane: Belle uses ellipses large enough to contain 90\% of signal
events, while \babar uses rectangular signal boxes
optimized to obtain the lowest expected upper limit in case of
no signal.  The Belle selection is significantly more effective in
suppressing the background, which is expected to be $0.01{-}0.07$
events in all channels but the three electron one, where 0.4 events
are expected due to Bhabha contamination. The
background in the signal region is estimated from the mass
distribution sidebands, assuming it is constant, using looser
selection criteria to get reasonable samples close to the signal
region. No details are given on how the extrapolation is done for the
full selection.  \babar expects $0.3{-}1.3$ background events, i.e.\
of order one, since the selection is optimized for the
best expected upper limit, at the risk of getting background events in
the signal region.  Belle observes no candidate signal events in
$535\invfb$ of data in all modes, and calculates upper limits
using Feldman and Cousin
ordering~\cite{Miyazaki:2007zw,Conrad:2002kn} in the range
$[2.0{-}4.1]\EE{-8}$, depending on the mode.  \babar observes from 0 to
2 events in $376\invfb$ of data, and calculates upper limits according
to Cousin and Highland prescription~\cite{Cousins:1991qz} with no
Feldman and Cousin ordering in the range $[4-8]\EE{-8}$.

\babar has published new results on tau LFV decays into a
lepton and a pseudoscalar meson
$\pi^0,\eta,\eta^\prime$~\cite{Aubert:2006cz}.
In these analyses both the $\eta\to\gamma\gamma$ and the $\eta\to
3\pi$ decay modes are used, and $\eta^\prime$
candidates decaying both to $\eta 2\pi$ and $\gamma 2\pi$ are
considered.
The expected background per channel is between 0.1 and 0.3 events.
Summing over all ten modes, 3.1 background events are expected, and 2
events are observed.

Belle has published improved results on tau LFV decays into a
lepton and a vector meson $V^0$~\cite{Nishio:2008zx}, with $V^0 =$
$\phi$, $\omega$, $K^{*0}$ or $\bar{K}^{*0}$, using 543\invfb of
data. No excess of signal events over the expected background is
observed, and upper limits on the branching fractions are obtained in the
range $(5.9{-}18)\EE{-8}$ at 90\%\,CL.

Belle has also reported on a LFV search for $\tau\to\ell f_0(980)$
followed by $f_0(980)\to\pi^+\pi^-$~\cite{Miyazaki:2008mw}: no events
are observed on a negligible expected background, obtaining 90\% CL
limits in the range $3.3{-}3.4\EE{-8}$ for the complete decay
chain. This channel is sensitive to Higgs mediated LFV processes like
$\tau\to\ell\eta$.

\babar reported in the past also on a less conventional search for LFV
in tau production ($e^+e^-\to\ell\tau$)~\cite{Aubert:2006uy}, finding
no signal.

The above results and additional B-factories LFV
results~\cite{%
  Aubert:2005tp,
  Aubert:2005ye,
  Aubert:2005wa,
  Aubert:2006uu,
  Yusa:2006qq,
  Miyazaki:2005ng,
  Miyazaki:2006sx,
  Miyazaki:2007jp,
  Hayasaka:2007vc
}
are summarized in Table~\ref{tab:lfvSearches}.

\begin{table}[tb]
  \begin{center}
  \caption{Summary of 90\%\,CL upper limits on tau LFV decays from the
    B-factories. An asterisk indicates a preliminary result. $h$ and
    $h^\prime$ denote a charged pion or kaon. Banerjee's combination of
    a subset of these channels is also included. For \babar, $V^0$
    includes just the $\omega$ vector meson.}
    \begin{tabular}{|l|cc|cc|}\hline
        \multirow{2}[0]{*}{Channel}
        & \multicolumn{2}{c}{Belle}
        & \multicolumn{2}{c|}{\babar}
        \\ \cline{2-5}

        & UL90 & Lumi 
        & UL90 & Lumi 
        \\

        & $(10^{-8})$ & (fb$^{-1}$)
        & $(10^{-8})$ & (fb$^{-1}$)
        \\ \hline

        $\mu\gamma$ & 5 & 535 & 6.8 & 232 \\
        $e\gamma$ & 12 & 535 & 11 & 232 \\
        $\mu\eta$ & 6.5 & 401 & 15 & 339 \\
        $\mu\eta^\prime$ & 13 & 401 & 13 & 339 \\
        $e\eta$ & 9.2 & 401 & 16 & 339 \\
        $e\eta^\prime$ & 16 & 401 & 24 & 339 \\
        $\mu\pi^0$ & 12 & 401 & 15 & 339 \\
        $e\pi^0$ & 8 & 401 & 13 & 339 \\
        $\ell\ell\ell$ & $2{-}4$ & 535 & $4{-}8$ & 376 \\
        $\ell h h^\prime$ & $20{-}160$ & 158 & $10{-}50$ & 221 \\ 
        $\ell V^0$ & $5.9{-}18$ & 543 &  $10{-}11$ & 384 \\

        $\mu K_S$ & 4.9 & 281 & & \\
        $e K_S$ & 5.6 & 281 & & \\
        $\mu f_0(980) \to \mu\pi^+\pi^-$ & 3.2\prelim & 671 & & \\
        $e f_0(980) \to e\pi^+\pi^-$ & 3.4\prelim & 671 & & \\
        
        $\Lambda\pi,\overline\Lambda\pi$ & $7.2{-}14$ & 154 & $5.8{-}5.9$\prelim & 237 \\
        $\Lambda K, \overline\Lambda K$ & & & $7.2{-}15$\prelim & 237 \\
        $\sigma_{\ell\tau}/\sigma_{\mu\mu}$ & & & $400{-}890$ & 211 \\
        \hline
        \multicolumn{5}{l}{($^*$ preliminary)}
      \end{tabular}
  \end{center}
  \label{tab:lfvSearches}
\end{table}

\subsection{Prospects}

While Belle plans to run until it will collect 1\,ab$^{-1}$ of data,
\babar has ended data-taking with a total collected luminosity of
about 530\invfb. In case of
no signal, the expected upper limits on the number of selected
signal events will improve depending of the amount of irreducible
background in each channel:
\begin{itemize}

\item
  when the expected background is large ($N_{\text{BKG}} \gg 1$), the
  expected upper limit is
  $N^{\text{UL}}_{90} \approx 1.64\sqrt{N_{\text{BKG}}}$;

\item
  when the expected background is small ($N_{\text{BKG}} \ll 1$),
  using~\cite{Cousins:1991qz} one gets $N^{\text{UL}}_{90} \approx
  2.4$.

\end{itemize}
Reducing the background below few events does not much improve the
expected limit if significant efficiency is lost in the process,
therefore optimized searches often enlarge the acceptance until
$N_{\text{BKG}} \approx 1$.  For the cleaner channels,
analyses can be optimized for an increased data sample to
keep $N_{\text{BKG}} \approx 1$ without loosing a significant part of
the signal efficiency: in this best case scenario, the expected upper
limits will scale as $N_{\text{BKG}}/{\cal L}$ i.e.\ as $1/{\cal L}$.
On the other hand, if no optimization is possible, just keeping the
current analyses will provide upper limits that scale as
$\sqrt{N_{\text{BKG}}}/{\cal L}$, i.e. as $1/\sqrt{{\cal L}}$.

Beyond the current B-factories facilities, there are proposals for
super B-factories~\cite{Bona:2007qt} that would permit a 100 fold increase
in the size of the tau pairs sample: this would allow probing tau LFV
decays at the $10^{-9}{-}10^{-10}$ level.

\section{$V_{us}$ measurement with tau decays}

$V_{us}$ measurements relying on kaon decays are limited by
theory uncertainties to a relative precision of $0.5\%{-}0.6\%$ (see
for instance ref.~\cite{Ambrosino:2008ct}).  On the other hand, the $V_{us}$
determination from the inclusive branching ratio of the tau to final
states with net strangeness equal to one ($\tau \to X_s \nu$) is
presently dominated from experimental uncertainties, and theory
uncertainties can be estimated as low as 0.23\%~\cite{Gamiz:2007qs}.

Recently, the \babar\ and Belle experiments have published
measurements that improve the former knowledge of the tau branching
fractions into
$\BR(\tau^- \to K^- \piz \nu_\tau)$~\cite{Aubert:2007jh},
$\BR(\tau^- \to \Kzb  \pi^- \nu_\tau)$~\cite{Epifanov:2007rf},
$\BR(\tau^- \to  K^-  \pi^- \pi^+ \nu_\tau)$~\cite{Aubert:2007mh}, and
$\BR(\tau^- \to  K^-  \phi \nu_\tau)$~\cite{Inami:2006vd,Aubert:2007mh}.
The information has been combined to obtain the most
up-to-date measurement of $\BR(\tau \to X_s
\nu) = $~\cite{Banerjee:2008hg}: $2.8447 \pm 0.0688$. When the
experimental measurement of \BFtautoknu is replaced with the
prediction based on the more precise $K_{\mu 2}$ measurement and
assuming $\tau-\mu$ universality~(\cite{Gamiz:2007qs}, $\BR(\tau \to
X_s \nu) = 2.8697 \pm 0.0680$~\cite{Banerjee:2008hg}.

Following ref.~\cite{Davier:2005xq}, assuming $\mu{-}e$ universality,
we use the tau branching fraction measurements to electron and muon to
obtain $\BR(\tau\to e \bar\nu_e \nu_\tau)^{\text{uni}} = \BR_e^{\text{uni}}
= (17.818 \pm 0.032)\%$, $R_{\tau,\text{hadrons}} = (1 - \BR_e -
\BR_\mu)/\BR_e^{\text{uni}} = 3.640 \pm 0.010$. Considering strange
and non-strange tau decays, $R_{\tau,\text{hadrons}} =
R_{\tau,\text{non-}s} + R_{\tau,s}$, where
$R_{\tau,s} = \BR(\tau\to X_s\nu) / \BR_e^{\text{uni}}$.
With the above we can measure:
\begin{eqnarray}
|V_{us}| &=& \sqrt{R_{\tau,s}/\left[\frac{R_{\tau,\text{non-}s}}{|V_{ud}|^2} - \delta R_{\tau,\text{th}}\right]},
\end{eqnarray}
where $\delta R_{\tau,th} = 0.240 \pm 0.032$~\cite{Gamiz:2006xx}. With
$V_{ud} = 0.97418(27)$~\cite{Amsler:2008zz}, we obtain $|V_{us}^\tau| =
0.2168 \pm 0.0028$. Unitarity predicts $|V_{us}^{\text{uni}}| =
\sqrt{1-|V_{ud}|^2} = 0.2259 \pm 0.0001$: the two values differ by
$3.06\sigma$.  On the other hand, $|V_{us}|$ obtained from kaon
measurements is about twice more precise and consistent with
unitarity~\cite{Ambrosino:2008ct}.

\section{Other results}

Several other results have been published with B-factories data, and
only a selection of the most recent ones is briefly mentioned in the
following.

Belle has measured the tau mass~\cite{Abe:2006vf} using $\tau \to
3\pi \nu_\tau$ with a pseudo-mass technique.  Although systematic
errors are larger than at threshold experiments, the tau mass can be
measured separately for oppositely charged tau leptons, permitting a
CPT test that found no violation.

Belle has reported preliminary measurements of several branching
fractions of the tau to final states including the $\eta$
meson~\cite{Inami:2008ar}. The measurements, including the invariant
mass distributions, are generally consistent with the
theory and adequately modeled by the Tauola
simulation~\cite{Was:2006my}.

\babar has measured $\BR{\tau\to 3\pi\eta\nu}$~\cite{Aubert:2008nj}
and has searched for the decay $\tau\to \eta^\prime(958)\pi\nu$, which
proceeds through a second-class current and is expected to be
forbidden in the limit of isospin symmetry. No evidence has been found
and a 90\%\,CL upper limit at $7.2\EE{-6}$ has been
obtained~\cite{Aubert:2008nj}.

Belle has precisely measured the branching fraction and the invariant
mass distribution of the decay $\tau^- \to \pi^- \pi^0 \nu_\tau$ using
72.2\invfb of data~\cite{Fujikawa:2008ma}. The unfolded invariant mass
spectrum is used to obtain the parameters for the $\rho(770)$,
$\rho^\prime(1450)$, and $\rho^{\prime\prime}(1700)$ meson resonances
and to estimate the hadronic ($2\pi$) contribution to the anomalous
magnetic moment of the muon ($a_{\mu}^{\pi\pi}$).

\section{Conclusions}

The B-factories have considerably advanced our knowledge on tau lepton
physics.  Searches for new physics effects have not found deviations
from the Standard Model expectations, and a number of upper limits
have been published to constrain new physics models.  The measurement
of $|V_{us}|$ with tau decays is promising because of the small
estimated error from theory, but significant work is needed to further
improve the precision on the most abundant Cabibbo-suppressed tau
decays. The first results from the B-Factories produce a low
$|V_{us}|$ value, which is about $3\sigma$ away from the unitarity
constraint derived from the precise experimental measurement of
$|V_{ud}|$.

\begin{acknowledgments}
I would like to thank the organizers of the HQL 2008 conference for
their hospitality and their patience in waiting for the delivery of
this contribution.
\end{acknowledgments}

\bigskip 
\ifflatbib

\else
\bibliography{%
  bibtex/pub-c-95-99,%
  bibtex/pub-d-00-04,%
  bibtex/pub-e-05-jun08,%
  bibtex/belle-2006-2008,%
  bibtex/tau-lepton,%
  bibtex/biblio-superb%
}
\fi

\end{document}